\documentclass{aa}  
\newcommand{\Gaia}{{\em Gaia}}
\usepackage{graphicx}
\usepackage{txfonts}

\begin{document}

   \title{Praesepe (NGC 2632) and its tidal tails.}

       \author{Siegfried R\"oser
         \inst{}
          \and
          Elena Schilbach
          \inst{}
          }
\institute
{Zentrum f\"ur Astronomie der Universit\"at
Heidelberg, Astronomisches Rechen-Institut, M\"{o}nchhofstra\ss{}e 12-14, 69120 Heidelberg, Germany\\
    \email{roeser@ari.uni-heidelberg.de, elena@ari.uni-heidelberg.de}}

   \date{Received 20 March 2019; accepted 16 April 2019.}

 
  \abstract
   {} 
{Within a 400~pc sphere around the Sun, we search for Praesepe's tidal tails in the Gaia DR2 dataset.}    
{We used a modified convergent-point method to find stars with space velocities close to the space velocity of the Praesepe cluster.} 
{We find a clear indication for the existence of Praesepe's tidal tails, both extending up to 165~pc from the centre of the cluster. A total of 1393 stars populate the cluster and its tails, giving a total mass of 794 M$_\odot$. We determined a tidal radius of 10.77~pc for the cluster and a tidal mass of 483 M$_\odot$. The corresponding half-mass radius is 4.8~pc. We also found clear indication for mass segregation in the cluster. The tidal tails outside 2 tidal radii are
populated by 389 stars. The total contamination of our sample by field stars lies between 50 to 100 stars or 3.6 to 7.2$\%$. We used an astrometrically and photometrically clean sub-sample of Gaia DR2 which makes our Praesepe sample incomplete beyond $M_G \approx$ 12.0 mag, which corresponds to about 0.25 M$_\odot$. A comparison with an N-body model of the cluster and its tails shows remarkably good coincidence. Three new white dwarfs are found in the tails.}  
   {}


   \maketitle
%
%
\section{Introduction}\label{intro}
Tidal tails around open clusters have been predicted about a decade ago by, e.g. \citet{2009A&A...495..807K} and \citet{2011A&A...536A..64E}. An introduction to this topic was given in our previous paper \citep{2019A&A...621L...2R}, hereafter called Paper I, and is not repeated here. The first detection of tidal tails of open clusters using the data from Gaia Data Release 2 \citep{2018A&A...616A...1G} were reported at the end of 2018 by \citet{2019A&A...621L...2R} and \citet{2019A&A...621L...3M} who revealed the tidal tails of the Hyades. This was followed by a number of publications where tidal tails have now been found around other open clusters.
\citet{2019arXiv190201404T} and  \citet{2019arXiv190207216F} recently reported about the detection of tidal tails of the Coma Berenices (Melotte 111) open cluster, and 
\citet{2019arXiv190104253Y} give hints for tidal tails from the very old, $\log \rm{t}$ = 9.33 \citep[MWSC,][]{2013A&A...558A..53K}, open cluster Ruprecht~147. So, tidal tails now seem to be a common feature of older open clusters and are revealed thanks to the high data quality of Gaia DR2.

NGC~2632 (Praesepe) is an old open cluster, $\log \rm{t}$ = 8.92 in MWSC, $\log \rm{t}$ = 8.85 according to \citet{2018A&A...616A..10G}. The latest major review of the stellar content of Praesepe \citep{2007AJ....134.2340K} dates back more than one decade.
Over an area of 300 $\deg^2$ on the sky, they found 1010 candidate members in Praesepe, and also estimated  the 
total mass to be 550 $\pm$ 40 M$_\odot$ and a tidal radius of 11.5 $\pm$ 0.3 pc. \citet{2013MNRAS.434.3236K} used 2MASS and SDSS photometry and proper motions from the PPMXL catalogue to determine the stellar content of Praesepe. They found a total cluster mass of about 630~M$_\odot$ and a 2-D half-mass radius of  3.90~pc. Using Gaia DR2 data,  \citet{2018A&A...616A..10G} and 
\citet{2018A&A...618A..93C} published membership lists of stars for quite a few open clusters in the Solar neighbourhood including Praesepe.
\citet{2018A&A...616A..10G} found 938 candidate members, and they also determined a mean parallax, mean proper motions, as well as age of, and reddening towards NGC~2632. For Praesepe, \citet{2018A&A...618A..93C} give membership probabilities for 719 stars as well as mean parallax and mean proper motions.  

This paper primarily aimed at the detection of the tidal tails of Praesepe, if they existed, but we also publish some characteristic astrophysical data of the cluster itself. The paper is structured as follows:
In Section \ref{detect} we describe the steps we took to find Praesepe's tidal tails. Section \ref{detect} is divided into subsections describing the cuts to obtain an astrometrically and photometrically clean sample, the separation of  over-densities in position and velocity space from the Galactic background, the identification of Praesepe and its tails, and the estimation of contamination and completeness. In Section \ref{CAMDs} we discuss the cluster and its tails and present some characteristic astrophysical parameters. A summary concludes the paper.
\begin{figure*}
\begin{minipage}[t]{0.490\textwidth}\vspace{0pt}
\begin{center}
\includegraphics[width=\textwidth]{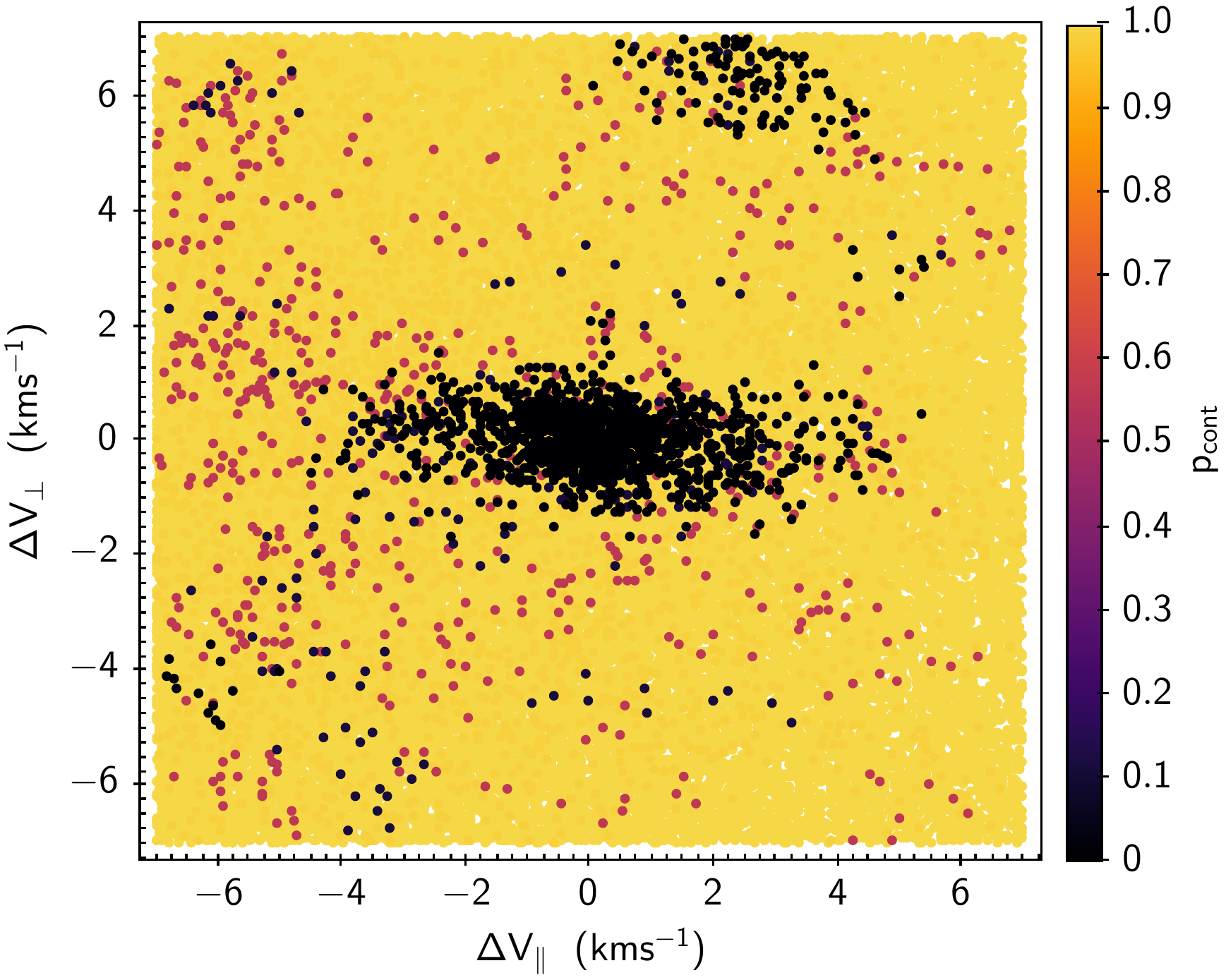}
\end{center}
\end{minipage}\hfill%
\begin{minipage}[t]{0.490\textwidth}\vspace{0pt}
\includegraphics[width=\textwidth]{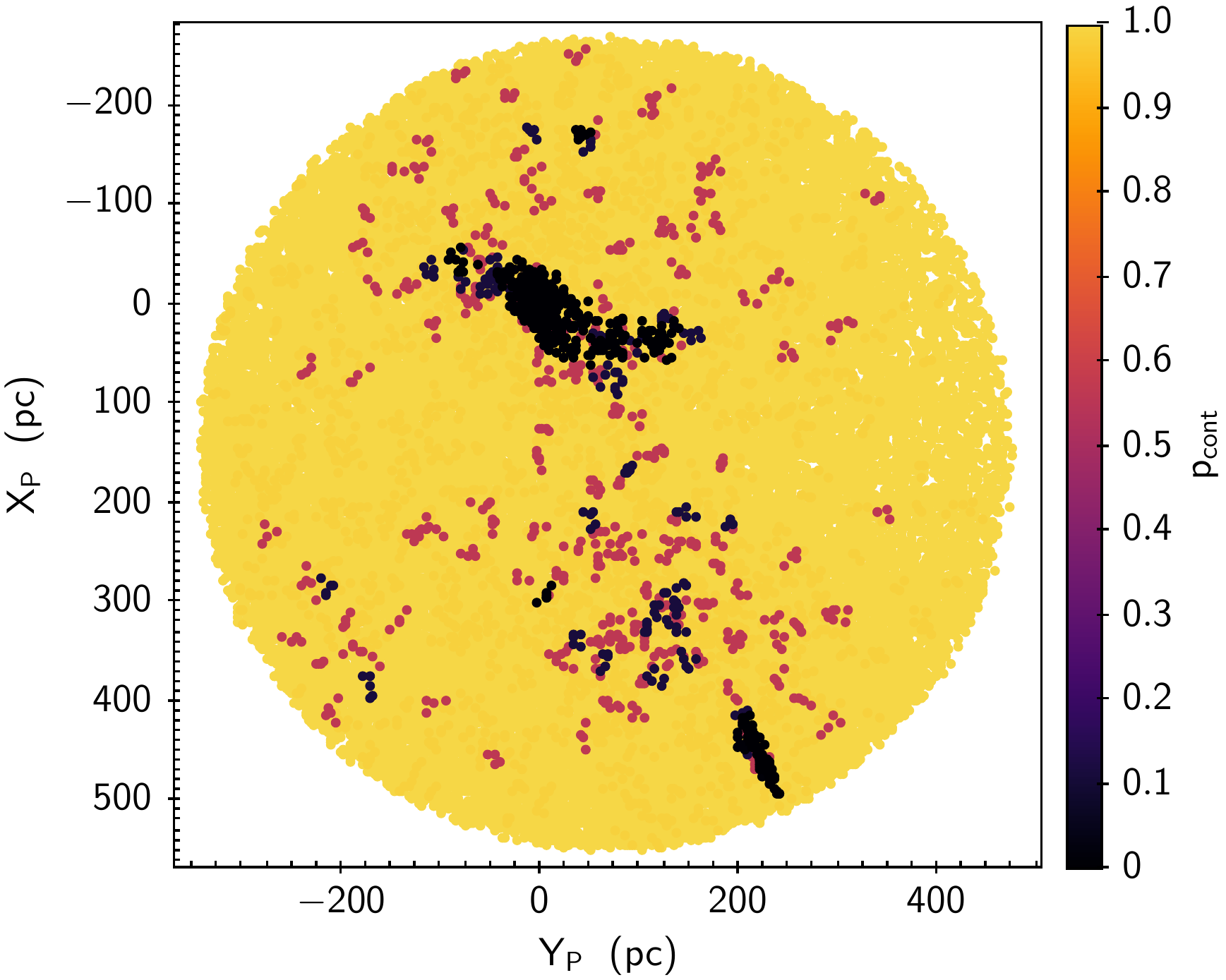}
\end{minipage}\\
      \caption{Left: Distribution of the stars from the basic sample (BS) in the tangential velocity plane $\Delta V_{\parallel}$, $\Delta V_{\bot}$ (see text for further explanation). The zero-point is the predicted value derived from the space motion of Praesepe. The $\Delta V_{\parallel}$-axis points in the direction of the convergent point. Stars are colour coded by their individual probability of contamination $p_{cont}$.  Right: the distribution of the same stars in the $Y_P,X_P$-plane. In this figure we use a Galactic Cartesian coordinate system with origin in the centre of the Praesepe cluster and not, as usual in the barycenter of the solar system. Praesepe and its tails are represented by the large dark structure at $X_P,Y_P$ = (0, 0). The over-density at $X_P,Y_P$ = (445pc, 220pc) is the open cluster Collinder 350.}
\label{Figure1}
\end{figure*} 
\section{Finding Praesepe's tidal tails}\label{detect}
The N-body simulations of the dynamical evolution of a Galactic open cluster  \citep{2009A&A...495..807K,2011A&A...536A..64E} literally were tailored to model the case of the Hyades cluster at an age of $\log \rm{t}$ = 8.8 \citep{2018ApJ...856...40M,2018ApJ...863...67G} and at about the galacto-centric distance of the Sun. As the galacto-centric distance and age of Praesepe are comparable to that of the Hyades, and as the clusters are also similar with regard to their stellar content, it is justified that we use the data set of the model from \citet{2009A&A...495..807K} as our model cluster to compare with the observations of Praesepe.

As a starting point we  determined the 6-D phase space coordinates of the centre of Praesepe on the basis of positions, proper motions, parallaxes, and radial velocities from Gaia DR2 and information on membership from \citet{2018A&A...616A..10G}. The coordinates are given in the solar system bary-centric Galactic Cartesian coordinate system where the $X$-axis points to the Galactic centre, the $Y$-axis in the direction of Galactic rotation, and $Z$-axis to the Galactic north pole. The corresponding velocity coordinates are $U, V, W$.
We found mean values
\begin{equation}
\begin{array}{lcl}
\vec{R_c}  =  (X_c,Y_c,Z_c) & = & (-141.75 , -69.35, 100.45)\,{\rm pc}, \\
\vec{V_c}  =  (U_c,V_c,W_c) & = & ( -43.17, -20.62, -9.60)\,{\rm km\,s^{-1}}.\label{COPO}
\end{array}
\end{equation}
These mean values are based upon 136 stars from \citet{2018A&A...616A..10G} with measured radial velocities. The standard errors of the mean in a position coordinate are 0.25~pc, and those of a velocity coordinate are 0.08~km~s$^{-1}$.
\subsection{Preparatory work}\label{prep}
From the Gaia DR2 dataset \citep{2018A&A...616A...1G} we extracted all
entries with a parallax greater than 2.4 mas, which gave 43,038,142 objects. 
For the further processing, we followed the procedures 
described in \citet{2018A&A...616A...2L}, Sect. 4.3 and Appendix C, Figs. C.1 and C.2, to obtain a stellar sample cleaned from possible artefacts with one small, but important change.
Instead of the  ``unit weight error"-cut (UWE) \citep[cf. Eq. C.1 in][]{2018A&A...616A...2L} we followed the new approach presented by Lindegren in the document \texttt{GAIA-C3-TN-LU-LL-124-01} which can be found on ESA's webpage\footnote{
\texttt{https://www.cosmos.esa.int/web/gaia/}\newline  \texttt{public-dpac-documents}}. In this concept UWE is scaled by a factor
depending on the magnitude and colour of the source. The result is a re-normalised UWE, or RUWE, which
needs only a single threshold of 1.4. Values for RUWE can be found at the service\footnote{\texttt{http://dc.zah.uni-heidelberg.de/tap}} in table \texttt{gaiaDR2light}.
All the other cuts are the same as in Paper I. We obtained an  
astrometrically and photometrically clean sample of 8,131,092 stars out of the original 43 million objects. The price for this cleaning is an incompleteness at the faint end of the magnitude distribution near $G$~=~18~mag and also near the parallax limit of 2.4~mas.
\subsection{Constraining the range in space and velocity}\label{CVspace}
In general, a cluster and its possible tidal tails reveal themselves as an over-density in position- and velocity-space.
In the ideal case, we need observed space velocities for each star in a sample to identify stars co-moving with the Praesepe cluster. However, accurate radial velocities are lacking for the vast majority of stars. Therefore we had to rely on criteria that are solely based on their tangential velocities. This implies that we may detect stars that are highly probable co-moving, although they need final confirmation when the radial velocities are measured. We followed the formalism of the convergent-point (CP) method as described in \citet{2009A&A...497..209V}, for instance, and transformed the Cartesian velocity vector of the cluster motion \vec{V_c} from Eq.\ref{COPO} to give predicted velocities $V_{\parallel pred}$ and $V_{\bot pred}$ parallel and perpendicular to the direction to the CP for each star depending on its position on the sky. We note that $V_{\bot pred} \equiv$  0. Also following \citet{2009A&A...497..209V}, we similarly transformed the measured (observed) tangential velocities for each star, $\kappa\,\mu_{\alpha*}/\varpi$ and $\kappa\,\mu_{\delta}/\varpi$ into $V_{\parallel obs}$ and $V_{\bot obs}$. Here $\varpi$ is the measured trigonometric parallax in Gaia DR2 and $\kappa=4.74047$ is the transformation factor from 1~mas~yr$^{-1}$ at 1~kpc to 1~km~s$^{-1}$. We set in the following 
$\Delta V_{\parallel}$ = $V_{\parallel obs}$ -  $V_{\parallel pred}$ and $\Delta V_{\bot}$ = $V_{\bot obs}$ -  $V_{\bot pred}$.
Note that  
the differences between the predicted and observed velocities ($\Delta V_{\parallel}$, $\Delta V_{\bot}$) will be equal to (0,0), when the space velocity of a star is identical to $\vec{V_c}$ in Eq.\ref{COPO}. 
We also determined the covariance matrix for the velocities $V_{\parallel obs}$ and $V_{\bot obs}$ via error propagation from
the covariance matrix of the $\mu_{\alpha*}, \mu_{\delta}$, and $\varpi$.

For further analysis we did not use the full sphere with $\varpi \leq$ 2.4~mas, but constrained the volume around Praesepe by  a cut in the z-plane as $|{Z - Z_c}| \leq 50$~pc where $Z_c$ is given in Eq. \ref{COPO}. This gives a volume of 4.867$\times 10^7~\rm{pc}^3$. We also introduced a restriction in the tangential velocity plane by requiring $|\Delta V_{\parallel}|\leq $7~km~s$^{-1}$ and  $|\Delta V_{\bot}| \leq$7~km~s$^{-1}$, a total area of 196~${(\rm{km~s}^{-1})}^2$. These cuts are ample compared to the predicted extent of the model tails, and they reduced our selection to a sub-sample containing 78,163 stars which populate this search volume. 

As we are interested in Praesepe and its possible tails, we also transform the spacial coordinates $X,Y,Z$ to a coordinate system  $X_P,Y_P,Z_P$ (the subscript $P$ refers to Praesepe) with origin at the centre of the Praesepe cluster, where the $X_P$-axis points to the Galactic centre, $Y_P$-axis to the direction of Galactic rotation and the $Z_P$-axis to the Galactic North Pole.
To achieve this we first went from the solar system bary-centric Galactic Cartesian coordinates $X,Y,Z$ to the Galactocentric cylindrical coordinate system $R,\theta,Z$ with $R_\odot$ = 8300~pc \citep{2013IAUS..289...29G} . Then we transformed back into the cluster-centric Galactic Cartesian system $X_P,Y_P,Z$, and finally $Z_P$ = $Z -Z_c$. Note that this transformation is not necessary for the way we determine over-densities, it only puts Praesepe into the centre of the volume we consider.
\subsection{A general method to find over-densities}\label{ITT} 
In Paper~I we cut a small rectangle in the tangential velocity plane where we suspected the signature of the tails, and then looked for over-densities in 3-D space. Here we follow a different, more general approach: we search simultaneously for over-densities in 5-D space. The main purpose is to separate the over-densities such as clusters or moving groups (signal) from the local Galactic background (noise).
In other words, for each star in the sub-sample of 78,163 stars we want to decide, if it belongs to the general Galactic field, or to an over-density significantly above the general background. We make the decision by counting neighbours of this star in a 5-D neighbourhood and compare this with the expected background counts where the background is modelled by a Poisson distribution. We make the following neighbourhood definition: a star $j$ with 5-D coordinates $(X_{P_{j}},Y_{P_{j}},Z_{P_{j}},\Delta V_{\parallel_{j}},\Delta V_{\bot_{j}})$  is a neighbour to star $i$ with coordinates $(X_{P_{i}},Y_{P_{i}},Z_{P_{i}},\Delta V_{\parallel_{i}},\Delta V_{\bot_{i}})$ if 
\begin{equation}
(X_{P_{i}}-X_{P_{j}})^2 + (Y_{P_{i}}-Y_{P_{j}})^2 +(Z_{P_{i}}-Z_{P_{j}})^2 \leq {r_{lim}}^2,
\end{equation} 
and
\begin{equation}
\frac{(\Delta V_{\parallel_{i}}-\Delta V_{\parallel_{j}})^2}{a^2} + \frac{(\Delta V_{\bot_{i}}-\Delta V_{\bot_{j}})^2}{b^2} \leq 1.
\end{equation}
The free parameters $a,b$ and ${r_{lim}}$ can be specified according to the goal of the study. 
We chose an elliptical shape for the velocity condition since the error ellipses in the tangential velocity plane can be eccentric. Indeed, in the case of the Praesepe sub-sample, $\sigma_{V_{\parallel obs}}$ and $\sigma_{V_{\bot obs}}$ are quite different; the mean $\sigma_{V_{\parallel obs}}$ is 0.79~km~s$^{-1}$, and the mean $\sigma_{V_{\bot obs}}$ is 0.12~km~s$^{-1}$. Considering also an intrinsic velocity dispersion in the cluster and tails, 
we chose 1.2~km~s$^{-1}$ as the semi-major axis $a$ of our selection ellipse and 0.5~km~s$^{-1}$ as the semi-minor axis $b$. With these velocity conditions we found, on average, 1.6$\times 10^{-5}$  stars for
each velocity ellipse and per 1~pc$^{-3}$.
Further we chose r$_{lim}$ = 15~pc to get, on average, 0.2 stars per 5-D neighbourhood volume.
Note that we can determine neighbourhood only for those stars, where the volumes specified for neighbourhood are fully contained in the subset given in Sec.~\ref{CVspace}. This holds for 54652 stars which we call the basic sample (BS).
\begin{table}
\caption{Distribution of the $k$-neighbourhoods of stars in the BS. For more information see text.}             
\label{table:1}      
\centering  
\begin{tabular}{|r|r|r|r|r|}
\hline
  \multicolumn{1}{|c|}{$k$} &
  \multicolumn{1}{c|}{$P_\lambda(k)$} &
  \multicolumn{1}{c|}{$N_{exp}(k)$} &
  \multicolumn{1}{c|}{$N_{obs}(k)$} &
  \multicolumn{1}{c|}{$p_{cont}$} \\
\hline
  0 & 0.761& 41587.5 & 41586 & 1.000\\
  1 & 0.208 & 11361.2 & 9952 & 1.142\\
  2 & 0.028 & 1551.8 & 1571  & 0.988\\
  3 & 0.003 & 141.3 & 252    & 0.561\\
  4 & 1.8E-4 & 9.6& 85       & 0.114\\
  5 & 9.6E-6 & 0.5 & 36      & 0.015\\
  6 & 4.4E-7 & 2.4E-2 & 42   & 5.7E-4\\
  7 & 1.7E-8 & 9.4E-4 & 20   & 4.7E-5\\
\hline\end{tabular}
\end{table}

When a star $i$ is surrounded by $k$ neighbours, we call this case a "$k$-neighbourhood". In the BS we found a total of 41586 zero-neighbourhoods and 9952 one-neighbourhoods, i.e.
the vast majority of stars (94.3\%) has less than 2 neighbours in the BS. With high probability, these are Galactic field stars (background noise). 
To determine an upper limit for the average density of the neighbours in the background we tentatively allowed all cases with up to 5 neighbours to be considered as background. These contain 97.9 per cent of the stars in the basic sample, which, on the other hand means that only the remaining about 2.1 per cent are signal. The average number of neighbours in the BS is 0.273 stars. So,
the distribution of neighbours in the background should follow a Poisson distribution with the mean $\lambda$ = 0.273.
We then calculated the quantity $N_{exp}(k) = P_\lambda(k) \times N_{BS}$ where $P_\lambda(k)$ is the probability mass function of the Poisson distribution representing the background. We give $P_\lambda(k)$ in the second and $N_{exp}(k)$ in the 3rd column of Table~\ref{table:1} for all
k $\leq$ 7. The fourth column in Table~\ref{table:1} gives the observed number of $k$-neighbourhoods $N_{obs}(k)$ in the BS. Comparing columns 3 and 4 we found that $N_{obs}(k)$ follows $N_{exp}(k)$ quite well for cases $k$=0,1,2. For increasing $k$ the number of observed $k$-neighbourhoods $N_{obs}(k)$ rapidly exceeds the prediction $N_{exp}(k)$.
If the case $k$=5 would belong to the background, the expected number of cases would be less than 1, observed are 36. So, we did certainly not underestimate the mean density $\lambda$ for the background distribution. In the case of Praesepe the highest value of $k$ is 405.
The ratio $p_{cont}$ = $N_{exp}(k)$/$N_{obs}(k)$  given in column 5 of Table~\ref{table:1} is the probability that a star in a $k$-neighbourhood belongs to the background distribution. Hence, for each star, we can interpret $p_{cont}$ as the degree how much it is contaminated by background.  According to the construction of these $k$-neighbourhoods a particular star from the BS can be found in the $k$-neighbourhood  of several other stars, and we give it the contamination $p_{cont}$ of the neighbourhood with the highest $k$, where this star is found.
\begin{figure*}
\begin{minipage}[t]{0.490\textwidth}\vspace{0pt}
\begin{center}
\includegraphics[width=\textwidth]{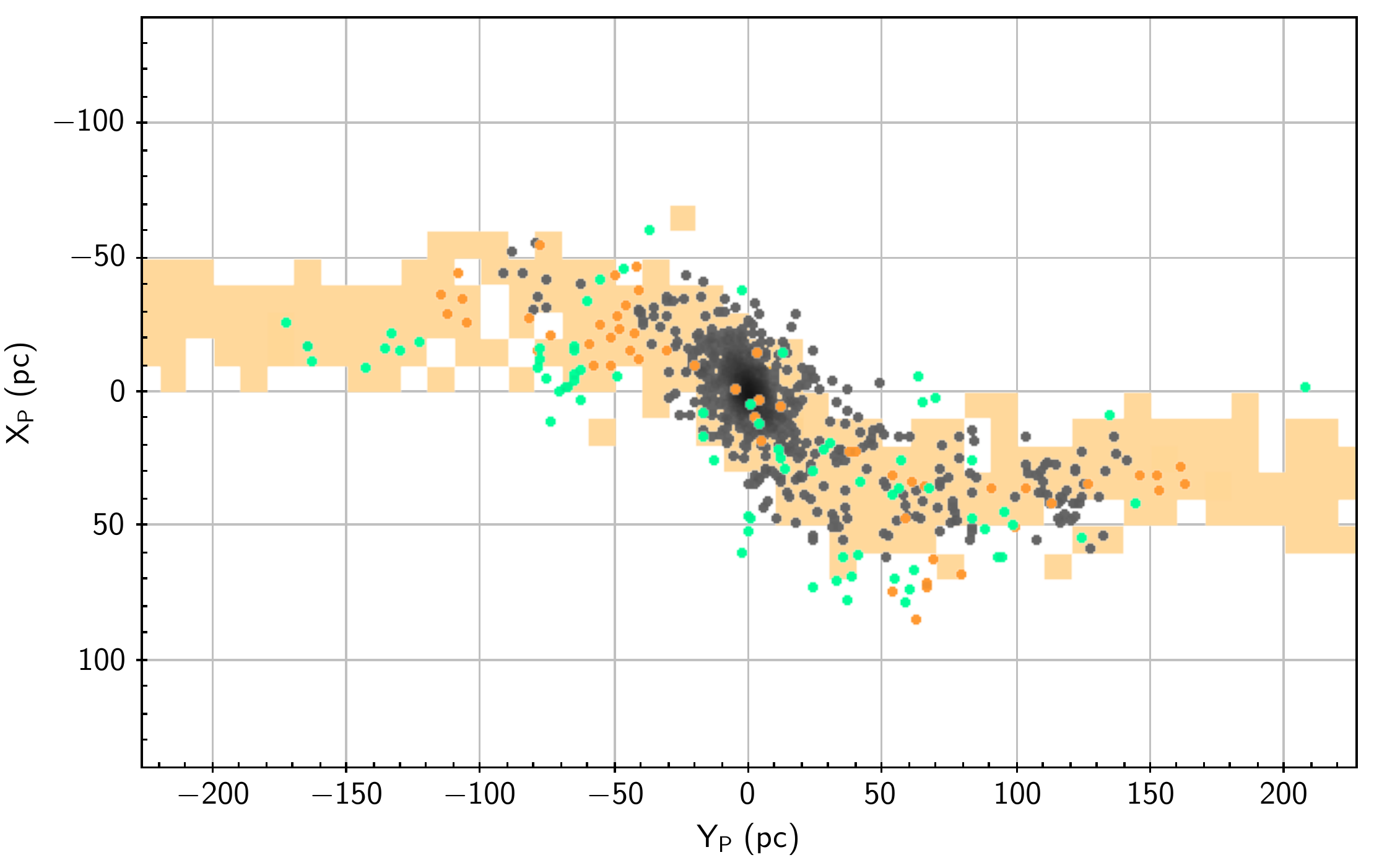}
\end{center}
\end{minipage}\hfill%
\begin{minipage}[t]{0.490\textwidth}\vspace{0pt}
\includegraphics[width=\textwidth]{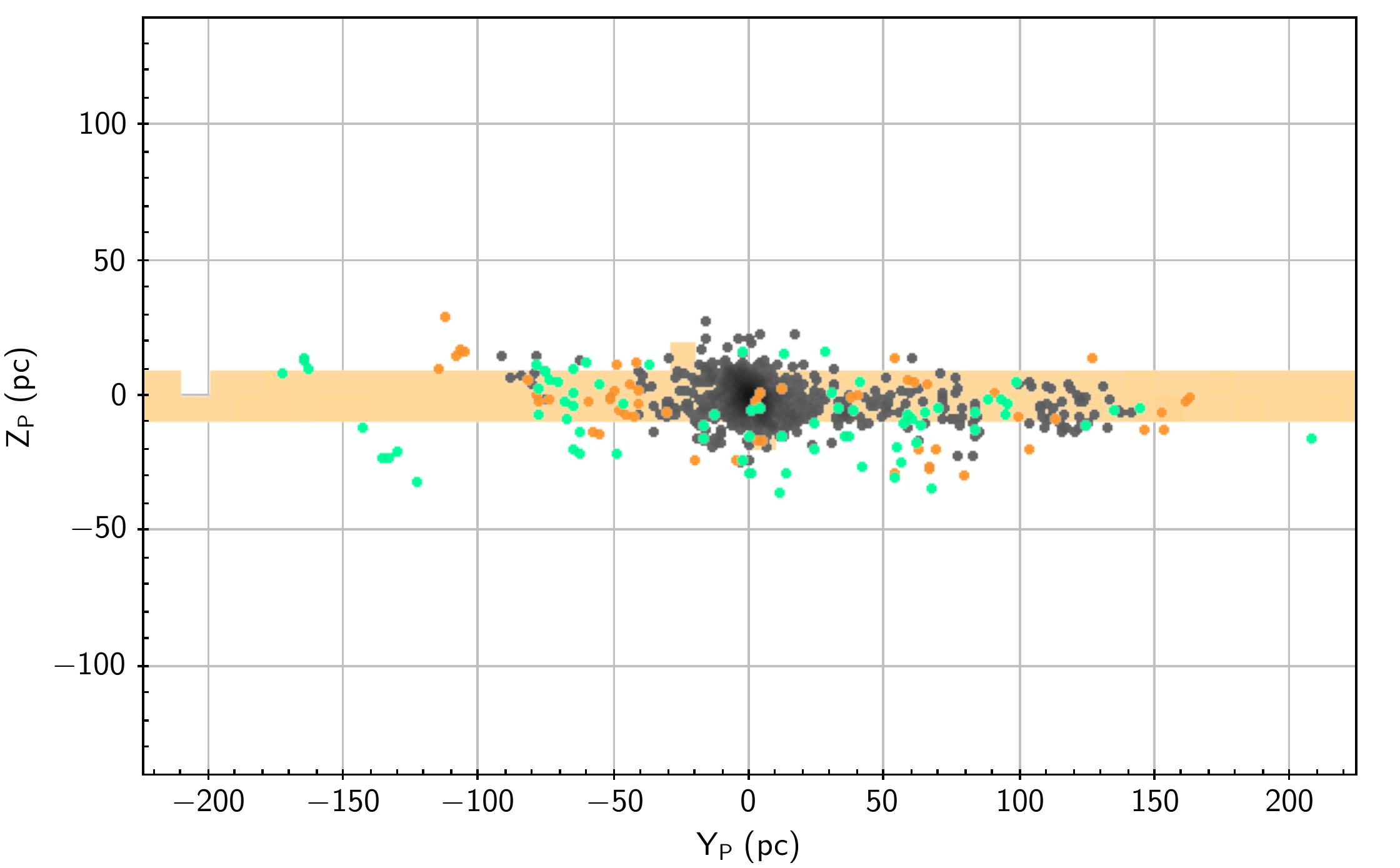}
\end{minipage}\\
      \caption{The Praesepe cluster and its tidal tails shown in the $Y_P,X_p$-plane (left) and the $Y_P,Z_p$-plane (right). Stars selected in Gaia DR2 are shown as dots colour-coded by their $p_{cont}$-values 
($p_{cont}$ <0.015 in grey, $p_{cont}$ $\approx$ 0.1 in orange and $p_{cont}$ $\approx$ 0.5 in light green). The location of the simulated tidal tails from the model by \citet{2009A&A...495..807K} is indicated as beige-coloured structures.}
\label{Figure2}
\end{figure*}  
In Fig.~\ref{Figure1} we show the distribution of all stars from the BS coded by their individual $p_{cont}$.
The left panel shows the distribution in the $\Delta V_{\parallel}$, $\Delta V_{\bot}$-plane, in the right panel we present the same set of stars, but in the $X_P,Y_P$-plane.
The background is filled by the vast majority (97.2\%) of stars from the BS having $p_{cont}$ $\approx$ 1 (cases $k$ = 0,1,2 in Table~\ref{table:1}), which
cover the whole area. Stars having  intermediate contamination (cases $k$ = 3,4) are clearly identifiable by their colour, and the remaining stars with $p_{cont} \approx 0$ are shown in black.
The huge signal centred at $\Delta V_{\parallel}$, $\Delta V_{\bot}$ = (0, 0) in the left panel and $X_P,Y_P$ = (0, 0) in the right panel 
represents Praesepe and its tails. A minor over-density at about $\Delta V_{\parallel}$, $\Delta V_{\bot}$ = (2.5, 6.3) and $X_P,Y_P$ = (445, 220) is the open cluster Collinder 350 and has nothing to do with Praesepe. In the following we considered only cases with $k$ $\geq$ 3 which we regard as signal. This yields 1543 stars or 2.8\% of the BS.

The selection of the parameters $(a,b)$ = (1.2 km~s$^{-1}$, 0.5 km~s$^{-1}$) in velocity space and ${r_{lim}}$ = 15~pc in positional space turned out to be a reasonable compromise to obtain a significant signal for Praesepe and its tails on one hand and a relatively small noise in the background on the other hand (see Fig.\ref{Figure1}). We tentatively kept $a$ and $b$ constant and increased ${r_{lim}}$ up to 25~pc. This resulted in a slight increase of the length of the tails together with a raise of spurious over-densities in the surrounding background created at velocities inconsistent with that of Praesepe. A similar effect occurred when we kept ${r_{lim}}$ constant at 15~pc and increased $(a,b)$ to (1.5 km~s$^{-1}$, 0.7 km~s$^{-1}$).

We used TOPCAT \citep{2005ASPC..347...29T} to extract the stars belonging to Praesepe and its tails out of the 1543 stars with $k$ $\geq$ 3. Three manual cuts, first in the $X_P,Y_P$-plane, then in the $\Delta V_{\parallel}$, $\Delta V_{\bot}$-plane and cut in the $X_P,Z_P$-plane were necessary to give the final number
(1393) of members (candidates) for Praesepe and its tails. Henceforth we call this set of 1393 stars the Praesepe Sample (PrS). By adding $p_{cont}$ of the stars  from the PrS we found a total contamination of 47 stars,
40 of which come from the stars in a 3-neighbourhood alone.

In Fig.~\ref{Figure2} we show the 1393 stars from Praesepe and its tails in the $(Y_P,X_P)$-plane and the $(Y_P,Z_P)$-plane. Stars 
coded in grey are 1266 stars with $p_{cont}$ <0.015, in orange 53 stars with $p_{cont}$ $\approx$ 0.1 and
in light green 74 stars  with  $p_{cont}$ $\approx$ 0.5. The tails extend up to 165 pc from the cluster centre. 
For comparison we also present the result of the N-body calculation for the Hyades cluster by \citet{2009A&A...495..807K}, which we also transformed into the $X_P,Y_P,Z_P$ coordinate system of Praesepe. The model stars are shown as a beige density plot in the background.
The coincidence between observations and model is remarkable, especially in the $(Y_P,X_P)$-plane. Such a good coincidence we already found in Paper I in the case of the Hyades. The present result could have been expected since the Hyades and Praesepe are very similar in age and in stellar content.
Among the stars in Fig. \ref{Figure2} which are slightly inconsistent with the model, we find that a majority of them has higher contamination (green and orange dots). A more rigorous cut in TOPCAT would have eliminated them, but it serves as an illustration how well our general selection method using the $k$-neighbourhood works. On the other hand, there are 5 stars with higher contamination in the trailing tail in Fig.~\ref{Figure2} that
are quite consistent with the model (left panel) and they are those farthest away from the cluster centre. At the same time these are the stars that are most distant from the Sun (up to 300~pc). The restrictions from Sec.~\ref{prep} applied to the full Gaia DR2 data set to get an astrometrically and photometrically clean sample creates some incompleteness here, especially at the faint end of the stellar sample. Note that incompleteness in signal leads to higher contamination by noise.

The internal contamination by field stars is given by summing up the values of $p_{cont}$. We sub-divided the PrS stars into 3 subgroups:
a) 725 stars within the tidal radius $r_t$ of 10.77~pc which is derived below, b) 279 stars between 1 and 2 tidal radii, and c) 389 stars outside 2 $r_t$ and in the tails. For these groups we found a contamination of 0.83 stars (group a), 0.69 stars (group b), and 46.1 stars or 11.8\% (group c). These values refer only to the contamination within our 5-D approach. At this stage it is also appropriate to estimate the possible field star contamination that comes from the fact that we did not consider radial velocities for estimating membership, since radial velocity data are only measured for a small sub-set of stars in the BS. Nevertheless, for 245 stars in the PrS  
radial velocities are available in Gaia DR2. Only 8 of them (3.3\%) have radial velocities inconsistent with Praesepe's space velocity.
The contamination increases from 1.2\% within the tidal radius to 10\% outside 2 $r_t$. Combining these 2 kinds of contamination we expect a total contamination between about 50 to 100 stars for our Praesepe Sample.
   \begin{figure}[hbtp]
   \centering
   \includegraphics[width=0.49\textwidth]{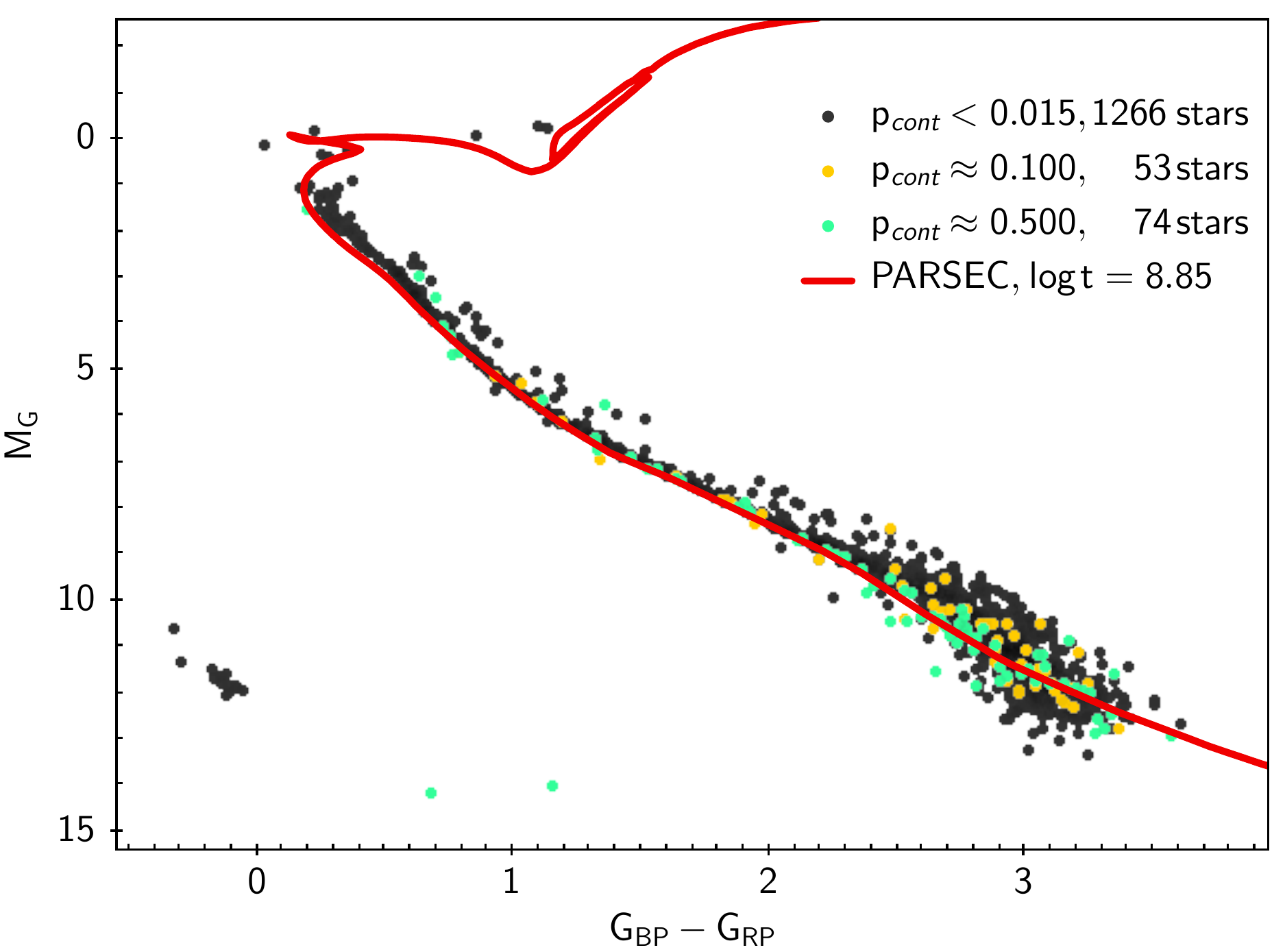}
      \caption{Colour-absolute-magnitude diagram (CAMD) $M_G$ versus $G_{BP} - G_{RP}$ of the 1393 stars of Praesepe and its tails. Stars are colour-coded by their $p_{cont}$. In addition we plot a PARSEC isochrone (Z = 0.02) with $\log \rm{t}$ = 8.85, and $E(B-V)$ = 0.027 mag.}
         \label{Figure3}
   \end{figure}
\section{Astrophysical properties of Praesepe}\label{CAMDs}
In Fig.~\ref{Figure3} we show the Colour-absolute-magnitude diagram (CAMD) $M_G$ versus $G_{BP} - G_{RP}$ of the 1393 stars of the Praesepe Sample. The stars are colour-coded according to their contamination value $p_{cont}$.  We also present a PARSEC-(version 1.2S)-isochrone \citep{2014MNRAS.444.2525C} with Z = 0.02 and $\log \rm{t}$ = 8.85, and $E(B-V)$ = 0.027 mag \citep{2018A&A...616A..10G}. The Gaia passbands from \citet{2018A&A...619A.180M} were used; we also applied the corrections to the Gaia $G$-magnitudes given in the same paper. We note here that the tails are spread over a wide part of space and, hence, the assumption of a unique reddening is not fully appropriate for tail stars. Nevertheless, over a wide range of $G_{BP} - G_{RP}$, the isochrone gives a reasonable fit to the observations except for the low-mass stars redder than about $G_{BP} - G_{RP}$ = 2.8 mag. Anyway, we used the isochrone shown in Fig.~\ref{Figure3} to roughly estimate individual masses via a mass-luminosity relation in the 3 Gaia photometric bands, neglecting binary issues. The incompleteness of our sample beyond $G$~$\approx$ 18 mag (see Sec.~\ref{prep}) creates incompleteness at $M_G$ $\approx$ 11.8 mag near the cluster centre, and brighter or fainter in the tails, depending on the individual parallax of the stars. Via our mass-luminosity relation this transfers to about 0.25 M$_\odot$.

The total mass of all stars from the PrS amounts to 794 M$_\odot$. 
We followed the recipes in  \citet{2011A&A...531A..92R} and used the individual masses to estimate the tidal radius of the Praesepe cluster.
In the galactic disk the sphere of influence of a
cluster is given by
\begin{equation}                 
    x_L = \left(\frac{\rm{G} M_t}{4A(A-B)}\right)^\frac{1}{3} 
  \label{tidal}       
\end{equation}
where $x_L$ is the distance of the Lagrangian points from the
centre, $M_t$ is the total mass inside a distance
$x_L$ from the centre, 
G = $4.3 \times 10^{-3} \rm{pc/M_\odot (km/s)^2}$ is the gravitational constant, $A$ and $B$ are 
Oort's constants at the position of the Sun. Here we use
 $A$ (15.3~$\rm{kms}^{-1}\rm{kpc}^{-1}$) and
$B$ (-11.9~$\rm{kms}^{-1} \rm{kpc}^{-1}$) from \citet{2017MNRAS.468L..63B}.
The distance $x_L$ is often referred to as the tidal radius $r_t$ of a cluster. The tidal radius $r_t$ separates,
in general, stars gravitationally bound to a cluster from those that are unbound. Adding up the individual masses from the centre outwards we found $r_t$~=~10.77~pc; 725 stars are within $r_t$ and give a tidal mass $M_t$ = 483~M$_\odot$ for Praesepe. Considering our incompleteness at about 0.25~M$_\odot$, this tidal radius is a lower bound. But, even if a total of 100~M$_\odot$ in low-mass stars with $m$ <  0.25~M$_\odot$ would be contained in the inner part of the cluster, the tidal radius would only rise to 11.5~pc. In models of star clusters the Lagrangian radius for 50\% of the cluster mass, plays an important role and is called half-mass radius $r_h$. Estimated from a tidal mass of 483 $\rm{M}_{\odot}$, the half-mass radius $r_h$ is 4.8 pc at the present state of evolution of Praesepe.

\citet{2007AJ....134.2340K} made an inventory of the Praesepe cluster about a decade ago. They used data from the SDSS, 2MASS, USNOB1.0, and UCAC-2.0 surveys, and analysed proper motions and photometry over 300 $\deg^2$ on the sky. They found 1010 stars in Praesepe being candidate members with probability > 80$\%$. These authors also estimated the total mass of the cluster to be 550 $\pm$ 40 M$_\odot$ and a tidal radius of 11.5 $\pm$ 0.3 pc. The small disagreement between our tidal radius and that from \citet{2007AJ....134.2340K} comes partly from their assumed values of $A$~=~14.4~$\rm{kms}^{-1}\rm{kpc}^{-1}$ and $B$ = -12.0 $\rm{kms}^{-1}\rm{kpc}^{-1}$. With the values from \citet{2017MNRAS.468L..63B}, \citet{2007AJ....134.2340K} would have got $r_t$ = 11.25~pc given that their derived total mass of 550~M$_\odot$ were inside 11.25~pc. Considering the fact that \citet{2007AJ....134.2340K} did not have available accurately measured trigonometric parallaxes, the coincidence of their tidal radius with ours is quite satisfactory. 
From their 1010 80$\%$-members we found 785 among our 1393 stars. From the remaining 225, only 77 passed the cuts in Sec.\ref{prep}, but were not selected as members when using our criteria. 

\begin{figure}[hbtp]
   \centering
   \includegraphics[width=0.49\textwidth]{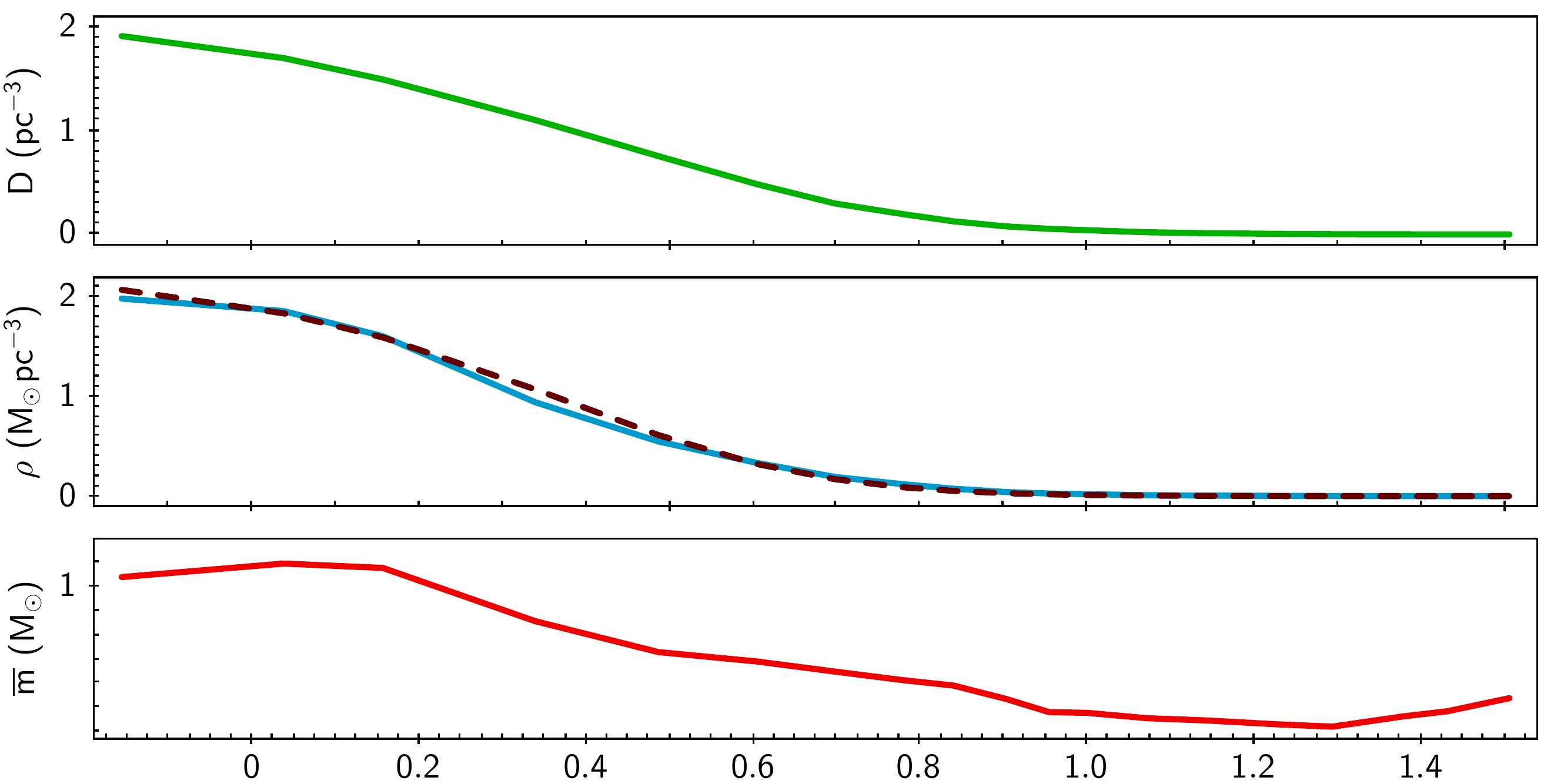}
      \caption{Number density $D$, mass density $\rho$, and average mass per star $\overline{m}$
      as a function of the logarithm of the distance from the centre ${r_{c}}$. A bin size of 2~pc in steps of 1~pc is used within 10~pc; beyond this the bin size is increased to 4~pc to get a better signal-to-noise ratio. 
      From top to bottom:
      the number density $D$ as a green curve,
      the observed mass density (blue curve) and a density distribution (black dashed line) from a Plummer model (Eq. \ref{pluden})
      with $M_t$ = 483 $\rm{M}_{\odot}$ and ${r_{co}}$ = 3.7~pc,
      the average mass per star  ($\rho/D$) as a red line. }
         \label{Figure4}
   \end{figure}

Using the individual masses we determine the radial density and mass profiles of the cluster. In Fig.~\ref{Figure4} we present the number density $D$, the mass density $\rho$ , and the
average mass per star as a function of the distance from the centre $r_{c}$.   
The bottom panel of Fig.~\ref{Figure4} reveals the mass segregation in Praesepe;
the average mass per star ($\rho/D$) decreases from 1.17 $\rm{M}_{\odot}$ close to the centre to about
0.46 $\rm{M}_{\odot}$ at the tidal radius $r_t$.  Again the apparent increase of the average mass per star towards larger distances from the centre is attributed to incompleteness at the low mass end. The mass density of Praesepe, shown in the middle part of Fig. \ref{Figure4}, is fitted
to a Plummer model \citep{1915MNRAS..76..107P}, where the mass density follows the equation
\begin{equation}
\rho({r_{c}}) = \frac{3M_t}{4\pi{r_{co}}^3}\frac{1}{[1+(r_c/{r_{co}})^2]^{5/2}}.
\label{pluden}
\end{equation}
Here $r_{co}$ is the so-called core radius of a cluster. Using the tidal mass of $M_t$ = 483 $\rm{M}_{\odot}$
the best fit to the observed density distribution (blue line) in Fig. \ref{Figure4} is
obtained with a core radius $r_{co}$ of 3.7 pc.
In the two upper panels of Fig.~\ref{Figure4} the core radius is seen as the distance
where the slope in the density and mass distributions changes.  
The corresponding Plummer model (dashed line in Fig.~\ref{Figure4}) shows good agreement
with observations. The ratio of the half-mass radius ${r_h}$ to ${r_{co}}$ in Praesepe is 1.297 which
is in remarkable coincidence with the theoretical ratio of 1.305 
for a Plummer model \citep[see][]{1975ApJ...200..339S}.

Fig.~\ref{Figure3} shows the presence of white dwarfs (WD) in Praesepe and its tails. In total, we find 13 white dwarfs with 9 of them inside the tidal radius of 10.77~pc. We disregard here the two objects with $M_G \approx 14$ mag and  $G_{BP} - G_{RP}$ near 1. Twelve out of 13 objects are known in SIMBAD \citep{2000A&AS..143....9W} as white dwarfs or blue proper motion stars. One star \object{Gaia DR2 62998983199228032} is not known as a WD in SIMBAD, but it fits perfectly to the loci of the other white dwarfs and has recently been rated as a member of Praesepe by \citet{2019MNRAS.483.3098S}.
All WDs by \citet{2019MNRAS.483.3098S} are confirmed by us with the exception of \object{WD 0840+205} which did not pass the quality cuts in Sec.~\ref{prep}. Three white dwarfs, \object{US 2088}, \object{SDSS J092713.51+475659.6} and \object{LAMOST J081656.19+204946.4} in the tidal tails have not been mentioned before as being related to Praesepe, but now we can confirm their provenance from Praesepe.
\subsection{A short comparison between Praesepe and the Hyades}\label{Praehya}
\begin{figure}[hbtp]
   \centering
   \includegraphics[width=0.49\textwidth]{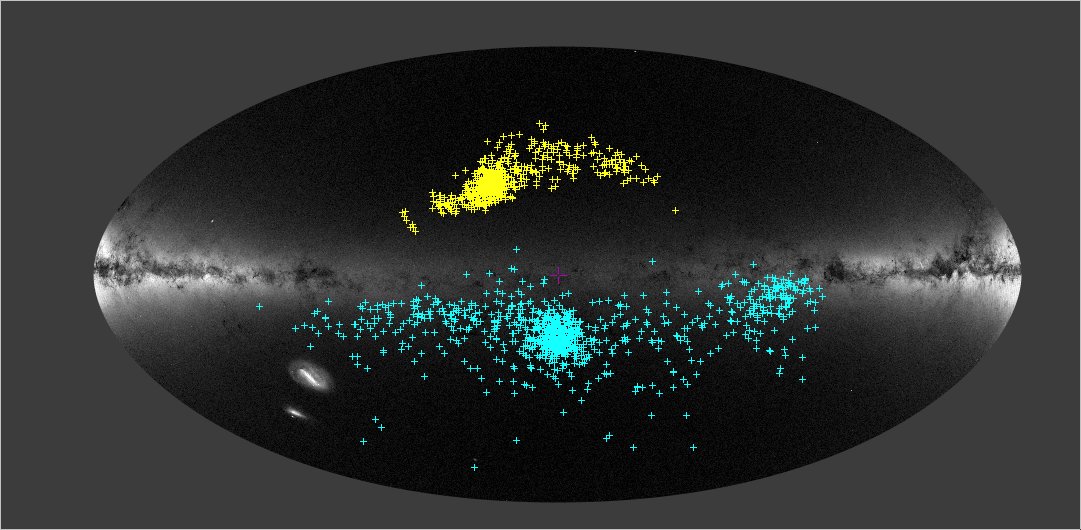}
      \caption{The tidal tails of Praesepe (yellow) and the Hyades (cyan) over-plotted to the Gaia DR2 sky using TOPCAT and Aladin facilities. }
         \label{Figure5}
   \end{figure}

In Fig.~\ref{Figure5} we show the location on the sky of Praesepe (yellow pluses) and the Hyades (cyan pluses) as well as the tidal tails of both clusters. The background is the grey-scale density map of Gaia DR2 (\copyright ESA/Gaia/DPAC). This composite image has been produced using the facilities of Aladin Desktop\footnote{Aladin Desktop is a product of CDS, Strasbourg, France}. 

A comparison with a theoretical model from N-body calculations \citep{2009A&A...495..807K} for an 650 Myr old cluster (Hyades) showed very good coincidence with regard to the location of the tidal tails of Praesepe, too. After the findings in Paper I, this is another very satisfactory confirmation for both theory and observations, in general. In detail, we observe for both clusters that the preceding tail is more pronounced than the trailing one, though we have good reasons to suspect that this is due to different causes. In the case of Praesepe the trailing tail stretches to larger distances from the Sun and probably suffers from the incompleteness of our sample at fainter stars. Both Hyades tails are located at about the same distance from the Sun but the trailing tail of the Hyades shows an inhomogeneous structure and may have been shredded by shocks or collisions in the past. Moreover, the proportion $N_{2rt}/N_{pt}$ of stars within the central part of the cluster, inside 2$r_t$ ($N_{2rt}$), to  stars in the preceding tail ($N_{pt}$) is considerably higher for Praesepe than for the Hyades, namely 5 and 2, respectively. This may indicate a later stage of dynamical evolution of the Hyades. They may be closer to dissolution than is Praesepe. However, this hypothesis must be supported by future investigations, e.g. N-body simulations.
\section{Summary}\label{summ}
Using the data from Gaia DR2, we searched for the presence of Praesepe's (NGC 2632) tidal tails in a sphere of some 400~pc radius around the Sun.
First, the Gaia DR2 data were cleaned according to the recipes given in  \citet{2018A&A...616A...2L} and described in Sec.~\ref{prep} to obtain an astrometrically and photometrically clean sample. 
Within this volume we cut a slice of $\pm$ 50 pc around the $Z_c$ coordinate of Praesepe and a primary window of  7~km~s$^{-1}$ by 7~km~s$^{-1}$ in velocity space around the expected tangential velocity of Praesepe members predicted by the convergent point method. We developed a method to find over-densities in a 5-D phase space with 3 spatial coordinates and 2 in velocity space. By modelling the Galactic background distribution via a Poisson distribution we were able to distinguish between over-densities and background noise. The over-density related to Praesepe is populated by 1393 stars. The background contamination in 5-D phase space amounts to 47 stars, and the estimate of contamination due to unknown discordant radial velocities gives about the same number. So the total contamination is estimated to lie between 50 and 100 stars or between 3.6 and 7.2\% of the Praesepe Sample. 
We used a 3-D ($M_G$, $M_{G_{BP}}, M_{G_{RP}}$) mass-luminosity relation to obtain individual masses for the stars from the PrS and found a total mass of 794 M$_\odot$. We determined the tidal radius ${r_t}$ of 10.77~pc and a tidal mass ${M_t}$ = 483 M$_\odot$ for Praesepe that is a lower limit since our sample gets incomplete at masses below 0.25 M$_\odot$. The half-mass radius ${r_h}$ is 4.8~pc. The shape of the cluster can be represented by a Plummer model with core radius ${r_{co}}$ = 3.7 pc. We observe a clear mass segregation of the cluster inside the tidal radius. The average mass per star drops from 1.17 $\rm{M}_{\odot}$ in the centre to about
0.46 $\rm{M}_{\odot}$ at the distance of the tidal radius ${r_t}$. 
We found 725 cluster members within one tidal radius; and a total of 1004 within two tidal radii. The remaining 389 stars from the Praesepe Sample populate two tidal tails. The tails extend up to about 165~pc on both sides along the $Y$-axis of Galactic rotation at the position of Praesepe. 

The locations of the tidal tails of Praesepe in 3-D space are in very good coincidence with a theoretical model for tidal tails \citep{2009A&A...495..807K} tailored to fit the 650 Myr old Hyades cluster. In detail, however, the ratio between the number of stars inside two tidal radii and outside (in the preceding tail) is much higher for Praesepe (factor 5) than for the Hyades (factor 2). This is a hint that the Hyades may already be in a later stage of evolution than is Praesepe.

We found 13 white dwarfs in our Praesepe Sample, 9 within one tidal radius and 4 outside. Three white dwarfs in the tail have not been assigned to Praesepe before, but this paper now identifies them as of Praesepe origin. We publish the data of all 1393 stars of the Praesepe Sample as on-line material.
\begin{acknowledgements}
We thank the referee, Carme Jordi, for her very constructive and useful comments that helped to improve the paper.
This study was supported by Sonderforschungsbereich SFB 881 ``The Milky Way
System" (subproject B5) of the German \emph{Deut\-sche For\-schungs\-ge\-mein\-schaft, DFG\/}. 
This research has made use of the SIMBAD database and the VizieR catalogue access tool, operated at CDS, Strasbourg, France.
This work has made use of data from the European Space Agency (ESA)
mission \Gaia\ (\url{https://www.cosmos.esa.int/gaia}), processed by
the \Gaia\ Data Processing and Analysis Consortium (DPAC,
\url{https://www.cosmos.esa.int/web/gaia/dpac/consortium}). Funding
for the DPAC has been provided by national institutions, in particular
the institutions participating in the \Gaia\ Multilateral Agreement.
\end{acknowledgements}
\bibliographystyle{aa}
\bibliography{mybib}
\end{document}